\documentclass[showpacs,floatfix,amsmath,amssymb, aps,twocolumn]{revtex4-1}

\usepackage{graphicx}
\usepackage{dcolumn}
\usepackage{bm}

\begin{document}

\preprint{APS/123-QED}

\title{Ultimate resolution of indefinite metamaterial flat lenses}

\author{Jessica B\'en\'edicto$^{1,2}$, Emmanuel Centeno$^{1,2}$, R\'emi Poll\`es$^{2,3}$,  Antoine Moreau$^{1,2}$}
\affiliation{$^1$Clermont Universit\'e, Universit\'e Blaise Pascal, Institut Pascal, BP 10448, F-63000 Clermont-Ferrand, France}
\affiliation{$^2$CNRS, UMR 6602, Institut Pascal, F-63177 Aubi\`ere, France}
\affiliation{$^3$Clermont Universit\'e, Universit\'e d'Auvergne, BP 10448, F-63000 Clermont-Ferrand, France}

\date{\today}

\begin{abstract}
We show that any metallo-dielectric multilayer with a hyperbolic dispersion relation can actually be characterized by a complex effective index. This refractive index, extracted from the complex Bloch band diagram, can be directly linked to the super-resolution of a flat lens made of this so-called indefinite medium. This allows for a systematic optimization of the lens design, leading to structures that are outperforming state-of-art flat lenses. We show that, even when fully taking absorption into account, our design provides super-resolved images for visible light up to a distance of one wavelength from the lens edge.
\end{abstract}
\maketitle

Since the seminal work of Sr J. B. Pendry, who proposed a perfect lens
of unlimited resolution, intensive efforts have been made to realize
reliable metamaterials that could make this concept effective
\cite{Pendry:2000p120}. Beating the optical diffraction limit requires
to conceive bulk metamaterials presenting both electric and magnetic
permittivities ($\epsilon$ and $\mu$) equal to -1. These slabs of -1
refractive index are however particularly difficult to realize at
optical frequencies since their building blocks (split rings
resonators, wires) are not sizable at the nanometer scale
\cite{Zheludev:2012hc}. An alternative approach has recently emerged
with feasible metal-dielectric multilayers, referred to as indefinite
metamaterials, which operate at ultraviolet frequencies
\cite{Drachev:2013gn}. These highly anisotropic media present
hyperbolic photonic dispersion surfaces in the k-space, which induces
enhanced optical properties \cite{Smith:2003bi, Sun:2013kz,
  Biehs:2013gj, Guo:2013cp, Yang:2012hg} Fig. 1. Consequently, both
propagating and evanescent waves emitted by a source decompose onto
propagating waves in the hyperbolic medium, so that the subwavelength
details are efficiently transported through the lens.  This principle
has been experimentally demonstrated in the far-field space with a
hyperlens that presents a spherical shape \cite{Smolyaninov:2007hr,
  liu, Narimanov:2007vm}. Planar indefinite hyperbolic metamaterials
have also been theoretically predicted to form subwavelength images
bonded at the output interface of the lens when the canalization
regime is reached \cite{Belov:2005, Belov:2006}. Others results have
moreover shown that light focalization is also possible in the
near-field and a lens equation for these hyperbolic lenses has been
derived \cite{Scalora:2007p97, Bloemer:2008tg, Mattiucci:2009wq,
  Benedicto}. Other progress towards negative index metamaterials have
been reported theoretically and demonstrated experimentally with
coupled plasmonic waveguide structures that resemble indefinite
metamaterials \cite{Verhagen:2010dy, Xu:2013cv}. The latter results
have in particular shown that these metamaterials behave as -1
effective index flat lenses that make images for ultraviolet light
\cite{Xu:2013cv}. This effective index has been evidenced by tracking
the refracted angles of a beam launched for various incident angles
through a stack of silver and TiO$_2$ thin films.
\begin{figure}
\centerline{\includegraphics[width=8.6cm]{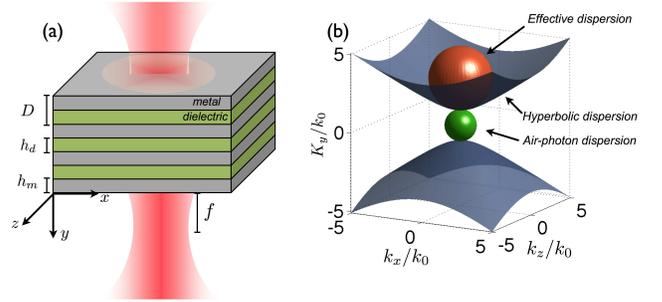}}
\caption{\label{fig:fig1} (color). (a) Principle of light focalization
  by indefinite metamaterials consisting of a multilayer of metal and
  dielectrics films. This hyperbolic lens focuses at the focal
  distance $f$. (b) Hyperbolic dispersion of an indefinite
  metamaterial, the green and red spheres represent respectively the
  dispersion relations of the air medium and of the effective medium.}
\end{figure}
According to the authors, subwavelength resolution has not been
achieved because of the optical losses, that are known to reduce the
efficiency of metamaterials lenses. However, very little is known so
far on the exact role of optical losses on the optimal optical
resolution of indefinite metamaterials lenses and, to our knowledge,
there is no theoretical model that allows a systematic optimization of
such systems. In this Letter, we develop a homogenization method,
called the paraxial homogenization method (PHM), that allows us to
define a complex effective index $\bar{n}=n+i\kappa$ for indefinite
metamaterials. Usual homogenization techniques start from the
constituents of the composite materials and deduce the effective index
required to determine the photon dispersion \cite{Vynck,
  Simovski:2009kf, Silveirinha:2007hh, Smith:2006p2995,
  Pendry:1999}. These bottom-up approaches, restricted to the long wavelength limit, lead to characterize indefinite metamaterials by an effective permittivity tensor \cite{Iorsh:2012da, Kidwai:2012fe}. The PHM is conversely a top-down approach
that enables to derive the effective index directly from the
dispersion relationship. As a consequence the PHM applies to both the
homogeneous and diffractive regimes and allows to calculate a
dispersive complex effective index. The design of optimal flat lenses
is thus guided by effective index and figure of merit $FOM=\left|
n/\kappa \right|$ maps obtained for various filling fractions of
metal-dielectric films. The optical resolution of indefinite
metamaterial lenses is directly derived from the complex band diagram
and is shown to be limited by the $FOM$ and the focal distance. These
results are successfully validated with a rigorous electromagnetic
calculations based on a modal method. Finally, indefinite
metamaterials with high $FOM$ are shown to efficiently transport the
evanescent waves to the rear face of the lens. This mechanism boosts the optical
resolution beyond the quarter-wavelength limit allowing to make
visible super-resolved images up to a focal distance of one
wavelength.

Consider a periodic set of $N$ slabs of silver and TiO$_2$ layers of thicknesses $h_{m}$ and $h_{d}$ and of complex relative permittivities $\bar{\epsilon}_m $ and $\bar{\epsilon}_d$, Fig. 1a. Silver films are placed at boundaries of the structure in order to benefit from surface plasmon polaritons resonances.  In the homogenization regime and for TM polarized electromagnetic waves propagating the y-direction, the effective permittivity tensor is characterized by the following diagonal elements:
\begin{equation}
\begin{array}{lcr}
\bar{\epsilon}_x= \bar{\epsilon}_m f_m+\bar{\epsilon}_d f_d \\
\bar{\epsilon}_y= \Big(\frac{f_m}{\bar{\epsilon}_m}+\frac{f_d}{\bar{\epsilon}_d}\Big)^{-1}
 \end{array} 
  \label{eq:epsilon}
\end{equation}
where the filling factors in metal and dielectric are respectively $f_m=h_m/D$ and $f_d=h_d/D$ with $D=h_{m}+h_{d}$  the lattice period. We denote by $L$ the total thickness of the multilayer. For convenience, these diagonal complex elements are written in the following form: $\bar{\epsilon}_x=\epsilon_x(1+i\sigma_x)$ and $\bar{\epsilon}_y=\epsilon_y(1+i\sigma_y)$. Provided that the conditions given in \cite{Benedicto} are satisfied, $\epsilon_y<0$ and $\epsilon_x>0$ and the hyperbolic relationship is $k_x^2/\bar{\epsilon}_y+\bar{k}_y^2/\bar{\epsilon}_x=(\omega/c)^2$. Here, $k_x$ refers to the conjugate coordinate of $x$ for Fourier Transform and reduces to a real number. We search how to replace this anisotropic medium by an equivalent homogeneous medium of complex effective index $\bar{n}$ that presents an isotropic photon dispersion $k_x^2+\bar{k}_y^2=(\bar{n}\omega/c)^2$, Fig. 1b. For a fixed frequency $\omega$, the paraxial homogenization method stems in deriving the complex  wavevector $\bar{k}_y$ in terms of a second order Taylor expansion:  $\bar{k}_y(k_x)=\bar{k}_y(0)-\bar{\gamma} k_x^2/{2k_0}$ where $\left. \bar{\gamma}=-k_0(\partial^2\bar{k}_y/\partial{k_x}^2)\right|_{k_x=0}$ and $k_0=2\pi/\lambda$.  For indefinite metamaterials, the complex curvature reads $\bar{\gamma}=\sqrt {\bar{\epsilon}_x} / \bar{\epsilon}_y$. This paraxial approximation allows us to replace this anisotropic medium by an isotropic one having the same complex curvature $\bar{\gamma}=1/\bar{n}$. When weak optical absorption is considered, the real part of the effective index and the figure of merit ($FOM$) are:
\begin{equation}
n=1/\Re(\bar{\gamma}),
\label{eq:indice}
\end{equation}
\begin{equation}
FOM=|\Re(\bar{\gamma}) / \Im(\bar{\gamma})|.
\label{eq:fom}
\end{equation}
In the homogenization regime, these effective parameters are analytically linked to the complex permittivity tensor elements by:
 \begin{equation}
n=-\frac{|\epsilon_y|}{\sqrt{\epsilon_x}}(1+\sigma_x\sigma_y/2)^{-1}, 
\label{eq:indice_hom}
\end{equation}
 \begin{equation}
FOM=\Big| \frac{2+\sigma_x\sigma_y}{2\sigma_y-\sigma_x} \Big|.
\label{eq:fom_hom}
\end{equation}
Equation \eqref{eq:indice_hom} shows that the effective index is governed by the ratio $-|\epsilon_y|/\sqrt{\epsilon_x}$ and decreases when the materials absorption increases. Despite these material losses, the $FOM$ is also seen to be maximal when $|2\sigma_y-\sigma_x|$ is minimal. As shown in \cite{Benedicto}, the dispersive properties of indefinite metamaterials strongly depends on the reduced frequency $D/\lambda$. As a consequence, the effective index and $FOM$ given by Eq. \eqref{eq:indice_hom} and \eqref{eq:fom_hom}  only apply when $D/\lambda<<1$. The PHM can however be extended in the diffractive regime, when the latter condition is not satisfied, by computing the curvature $\bar{\gamma}$ from the complex band structure \cite{Bergmair:2006cw}. In this way, the real and imaginary parts of the complex curvature are respectively $\left. \Re(\gamma)=-k_0(\partial^2\Re(\bar{K}_y)/\partial{k_x}^2)\right|_{k_x=0}$ and $\left. \Im(\gamma)=-k_0(\partial^2\Im(\bar{K}_y)/\partial{k_x}^2)\right|_{k_x=0}$ where $\bar{K}_y$ is the complex Bloch wavevector derived from dispersion equation \cite{Bergmair:2006cw, Polles:2011p5453}. 
\begin{figure}
\centerline{\includegraphics[width=8.6cm]{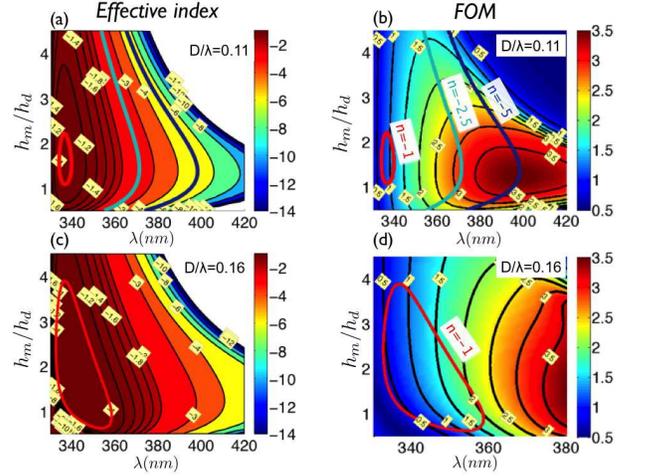}}
\caption{\label{fig:fig2} (color). Effective index and $FOM$ with respect the wavelength and the metal-dielectric ratio for: $D/\lambda=0.11$ (a-b) and $D/\lambda=0.16$ (c-d). Equi-index curves of -1, -2.5 and -5 effective indices (red, green and blue curves) are superposed in the $FOM$ maps.}
\end{figure}
This principle is depicted on Fig. 2 where  effective index and $FOM$ are plotted for two reduced frequencies $D/\lambda=0.11$ and $D/\lambda=0.16$  according to equations \eqref{eq:indice} and \eqref{eq:fom}. These maps enable to compute the indefinite metamaterials effective properties for various metal-dielectric filling factors and wavelengths by considering the complex permittivities of silver and TiO$_2$ \cite{TIO2, Palik}. The effective index is shown to range from -1 to about -10 for wavelengths between $330\,nm$ and $420\,nm$. The effective index is in addition red-shifted when the reduced frequency increases, leading to a wider -1 equi-index curve for $D/\lambda=0.16$. The $FOM$ is shown to be optimal around $\lambda=400 \,nm$ and when $h_m/h_d$ is close to $1.3$. From these maps, we conclude that the best flat lens made with indefinite metamaterials with a -1 refractive index is obtained for a $FOM=2$ when the following parameters hold:  $D/\lambda=0.16$, $\lambda=358\,nm$, $h_m/h_d=1.07$.  These parameters lead to silver and TiO$_2$ thicknesses of $h_m=28\,nm$ and $h_d=30\,nm$ that are very close to the parameters used by T. Xu \textit{et al.} in their experimental demonstration \cite{Xu:2013cv}. Looking for lower optical losses leads to a lens with a $FOM$ as high as 4, obtained for $D/\lambda=0.11$ at $\lambda=398\,nm$ and $h_m/h_d=1.28$. But in that case, the hyperbolic flat lens is characterized by an effective index of -5.

In order to compare the optical efficiency of these -1 and -5 effective index metamaterials, respectively referred to as $IM^{(-1)}$ and $IM^{(-5)}$, we derive a model that directely links their $FOM$ to the optical resolution. We assume an incident Gaussian beam of waist $W$ that impinges the input interface. After propagating through the lens, whose thickness is $L$, the transmitted signal calculated at a distance $y$ far away from the output interface reads:
\begin{equation}
U(x,y)=\int_{-\infty}^{\infty}  A(k_x) T(k_x) t_0(k_x,y) e^{i k_x x} dk_x
\label{eq:propagation}
\end{equation}
where $t_0(k_x,y)=e^{iy\sqrt{k_0^2-k_x^2}}$ is the transfer function in the output air medium and $A(k_x)=W_0/(2\sqrt \pi)\exp(-(k_x W/2)^2)$ is the spectrum of the incident Gaussian beam in the k-space. The multilayer transmission coefficient, $T(k_x)$, splits into a singular part $T^s(k_x)=\sum_{p} t_p/(k_x-\bar{k}_x^{(p)})$ where $\bar{k}_x^{(p)}$ and $t_p$ are the poles and the residues associated to optical resonances and a regular part $T^r(k_x)$. Since this holomorphic function accounts for both the phase and the optical absorption accumulated by the signal when propagating through the metamaterial slab,  we assume that it is driven by the complex Bloch wavevector: $T^r(k_x)=e^{i\bar{K}_y(k_x)L}$. The use of the paraxial expansion of $\bar{K}_y(k_x)$ allows $T^r$ to be expressed in terms of the effective index and $FOM$:
\begin{equation}
 T^r (k_x)= e^{i\bar{K}_y(0)L}  e^{-\frac{k_x^2}{2k_0} \frac{L}{|n| FOM}} e^{i\frac{k_x^2}{2k_0} \frac{L}{n} }
 \label{eq:Tr}
 \end{equation}
An image is obtained when the focalization power provided by the lens cancels out the optical diffraction in air i.e. when the total phase vanishes at the focal distance $f$. As shown in \cite{Benedicto}, the paraxial approximation applied to $t_{air}$ leads to $f=-L/n$ where $n$ is the effective index defined in Eq. \eqref{eq:indice} and $L$ is the lens thichness. Finally, the focalized beam is the sum of a resonant field $U^s(x,f)$ (linked to ($T^s(k_x)$) and a regular field $U^r(x,f)$ given by:
 \begin{equation}
U^r(x,f)=\int_{-\infty}^{\infty}  A(k_x) |T^r(k_x)| |t_0(k_x,f)| e^{i k_x x} dk_x
\label{eq:propagation}
\end{equation}
Perfect resolution  implies that the transfer function product  $|T^r(k_x)| |t_0(k_x,f)|=1$ where $|t_0(k_x,f)|=e^{-f\sqrt{k_x^2-k_0^2}}$.  However, Eq. \eqref{eq:res}, derived from Eq. \eqref{eq:Tr}, shows that evanescent waves  (for $k_x>k_0$) are rather reduced  than amplified inside the lens, contrarily to what would be expected for a perfect lens \cite{Pendry:2000p120} : 
\begin{equation}
 |T^r(k_x)|=   e^{-\Im(\bar{K}_y(0))L} e^{-\frac{k_x^2}{2k_0} \frac{L}{|n| FOM}} 
\label{eq:res}
\end{equation}
\begin{figure}
\centerline{\includegraphics[width=8.6cm]{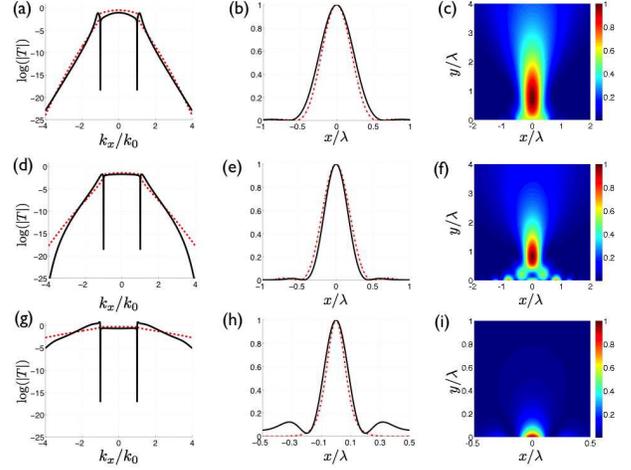}}
\caption{\label{fig:fig3} (color). Logarithm of  the transfer functions, intensity profiles obtained at the focal distance and maps of focalized beams. The solid and dashed curves correspond respectively to the exact transfer function and the paraxial one $T^p$. Figures (a-c) and (d-f) are obtained for $f=\lambda/2$ with $IM^{-1}$ and  $IM^{-5}$ respectively. Figures (g-i) correspond to $IM^{-5}$  for $f=\lambda/17$.}
\end{figure} 
As a consequence of this description, the optimal transport of subwavelength details requires a large $FOM$ but it is limited by the focal distance. We apply these results to the optimized indefinite metamaterials $IM^{(-1)}$ and $IM^{(-5)}$ of respective effective indices -1 and -5. The exact  transfer function computed with a rigorous modal transfer matrix method \cite{Krayzel:2010jg} is compared to the total transfer function $T^p=|T^r(k_x)| |t_0(k_x,f)|$ derived from the  paraxial approximation of Eq. \eqref{eq:res}, Fig. 3.  For that purpose, the focal distance is initially kept to a half of wavelength ($f=\lambda/2$) by using the convenient number of lattice periods and a Gaussian beam of   subwavelength waist $W_0=\lambda/20$ is launched toward the input lens interface.  It can be seen on Fig. 3a and 3d that the paraxial transfer function $T^p$ agrees well with the rigorous one and that its slope is reduced when the $FOM$ increases. The sharp oscillations observed around $k_x=\pm 1.2k_0$  correspond to poles of the singular part. These resonances, associated to surface plasmon polaritons (SPPs) like resonances bounded at the rear lens interface, are however only activated for $IM^{-5}$, Fig. 3c and 3f. Despite the fact that these surface modes are not caught by paraxial transfer function, the derived intensity profiles match almost perfectly with the rigorous computations. This demonstrates the validity of the PHM, Fig. 3b, 3e and 3h. Finally, the full width at half maximum (FWHM) calculated for this focal distance shows that subwavelength resolution of $\lambda/2.7$ is achieved for $IM^{(-5)}$ (with a $FOM$ of 4) while the optical resolution of $IM^{(-1)}$ is limited to $\lambda/1.9$ since its $FOM=2$. A high $FOM$ allows to efficiently transport both the propagating and evanescent waves, the latter particularly boosting the resolution. This mechanism enhances the optical resolution by optimizing the regular field $U^r(x,f)$ and increases the contribution of the singular field $U^s(x,f)$. A maximal resolution of $\lambda/6$ is reached for $IM^{(-5)}$ when the image is focused at the output lens interface, Fig 3(g-h-i). Beyond this canalization regime, super-resolution persists as long as the focal distance is smaller than $\lambda$, showing the crucial impact of high $FOM$ and SPP-like resonances, Fig. 4. The lower $FOM$ of the $IM^{-1}$ device only enables super-resolution for a short focal distance of $\lambda/3$ which corresponds to a multilayer of 3 lattice periods. Beyond this distance, subwavelength details carried by high-k components are irreversibly lost because of the optical losses.
\begin{figure}
\centerline{\includegraphics[width=8cm]{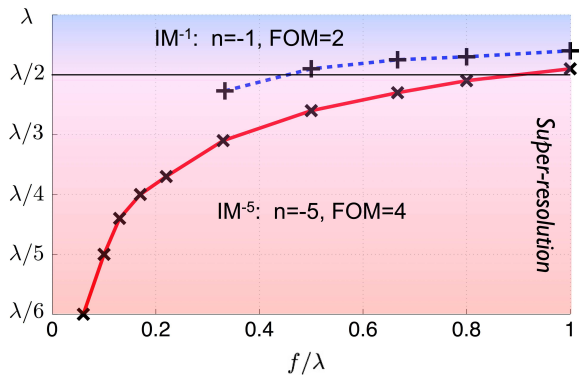}}
\caption{\label{fig:fig4} (color) Optical resolution with respect to the focal distance in unit of $\lambda$ for $IM^{-1}$ (dashed curve) and $IM^{-5}$ (solid curve). Super-resolution (red area) is obtained with $IM^{-5}$ as long as the focal distance is smaller than one wavelength. Each points are obtained with the rigorous transfer matrix method \cite{Krayzel:2010jg}. }
\end{figure} 

In conclusion, we have demonstrated that indefinite metamaterials characterized by an anisotropic permittivity tensor are equivalent to an isotropic homogeneous metamaterials of complex negative effective index. This effective refractive index and its associated absorption constant are extracted from the complex Bloch diagrams by the use of a paraxial homogenization method that applies in both the homogenization and diffractive regimes. This very general approach, that could also be extended to the design of optimized hyperlenses, gives maps of effective index and $FOM$ for various metal-dielectric compositions and wavelengths. We have in addition developed a semi-analytical theory that links the $FOM$ to the optical resolution of such lenses. These results, validated by rigorous ab-initio electromagnetic computations, show that  indefinite metamaterials of large $FOM$ are required to efficiently transport high-k components of the signal.  Subwavelength resolution is then attributed to a SPP-like resonance assisted mechanism rather than the \textit{amplification} of evanescent waves theorized for left-handed metamaterials. Finally, we show that a feasible hyperbolic lens of -5 effective index can outperform the optimal lens of -1 effective index  by allowing visible light focalization (at $\lambda=400 \,nm$) together with super-resolution for focal distances as large than one wavelength.

We acknowledge K. Vynck and  C. Sauvan for discussions.

\end{document}